\documentclass[fleqn,twoside]{article}
\usepackage{espcrc2}

\usepackage{graphicx}

\def\bc{\begin{center}}
\def\ec{\end{center}}
\def\beq{\begin{equation}}
\def\eeq{\end{equation}}
\def\bs{\begin{slide}}
\def\es{\end{slide}}
\newcommand{\bmath}{\begin{displaymath}}
\newcommand{\emath}{\end{displaymath}}
\newcommand{\beqn}{\begin{eqnarray}}
\newcommand{\eeqn}{\end{eqnarray}}
\newcommand{\beqns}{\begin{eqnarray*}}
\newcommand{\eeqns}{\end{eqnarray*}}
\newcommand{\ba}{\begin{array}{c}} 
\newcommand{\bat}{\begin{array}{cc}} 
\newcommand{\ea}{\end{array}} 
\newcommand{\lef}{(1-\gamma_5)}

\newcommand{\lsim}{\stackrel{<}{_\sim}}
\newcommand{\gsim}{\stackrel{>}{_\sim}}
\newcommand{\gev}{\, \mbox{GeV}}
\newcommand{\Frac}[2]{\frac{\displaystyle #1}{\displaystyle #2}}

\title{Hadronic decays of the tau lepton~: Theoretical overview 
\thanks{IFIC/04$-$69 report. Talk given at the International Workshop on Tau
Lepton Physics, TAU04 (14-17 September 2004), Nara (Japan).}}

\author{J. Portol\'es \address[IFIC]{Instituto de F\'{\i}sica Corpuscular, IFIC,
                                     CSIC-Universitat de Val\`encia, \\
                                     Edifici d'Instituts de Paterna,
                                     Apt. Correus 22085, E-46071 Val\`encia, Spain}}
       
\begin{document}

\begin{abstract}
Exclusive hadronic decays of the tau lepton provide an excellent framework to study
the hadronization of QCD currents in a non-perturbative energy region populated by
many resonances. I give a short review both on the main theoretical tools employed to 
analyse experimental data and on how Theory compares with Experiment.
\end{abstract}

\maketitle
\section{Introduction}
Strong interactions span their effects in all electroweak generated processes whenever
hadrons are involved. 
In spite of the success of Quantum Chromodynamics in the description and understanding
of the strong interaction since the very early stages of its applications, it soon became
clear that, due to one of its most relevant characteristics, asymptotic freedom, the 
study of the low energy processes involving the lower part of the hadronic spectrum
(typically $E \lsim 2 \gev$) would be not possible with a strong interaction theory
written in terms of quarks as dynamical degrees of freedom.
Therefore while in the perturbative domain QCD provides a well defined and
successful framework to describe the strong interaction, in the very low or intermediate
energy region, heavily populated by resonances, one has to 
resort to procedures that intend to collect the main features of the
underlying theory. 
Accordingly several main paths have
been followed in order to handle hadron dynamics in this energy regime~: 
\vspace*{0.2cm} \\
\underline{\em Effective field theories}
\vspace*{0.2cm} \\ \hspace*{0.3cm} In the early eighties, and
relying in a very fruitful heritage from the pre-QCD era, it was observed \cite{chiral,chiral1}
that the chiral symmetry of massless QCD could be used to construct a strong interaction
field theory, intended to be dual to QCD, in terms of the lightest $SU(3)$ octet of pseudoscalar mesons in the role of pseudo-Goldstone bosons associated to the spontaneous breaking of that symmetry. This construction, known as Chiral Perturbation Theory ($\chi PT$), has been
very much useful in the study of strong interaction effects at very low energy, where the
theory has its domain, mainly $E \ll M_{\rho}$ (being $M_{\rho}$ the mass of the $\rho (770)$,
the lightest hadron not included in the theory as an active field). Its success has pervaded
hadron physics in the last two decades, bringing to the main front the concept of
effective field theory as a powerful tool to handle the non-perturbative regime of QCD. 
An effective field theory tries to embody the main features of the fundamental theory in order
to handle the latter in a specific energy regime where is, whether more inconvenient or just
impossible, to apply it \cite{eft}. The example of $\chi PT$, together with other developments that we will comment later, gave support to further extensions in the spectrum of hadrons described
by the theory, like the lightest $SU(3)$ octets of vector, axial-vector, scalar and pseudoscalar
resonances, in the Resonance Chiral Theory ($R \chi T$) that provides a tentative
framework to study 
the energy region given by $M_{\rho} \lsim E \lsim 1.5 \gev$.
\vspace*{0.2cm} \\
\underline{\em Modelizations of phenomenological Lagrangians} \vspace*{0.2cm} \\ \hspace*{0.3cm}
As a sidetrack of effective field theories many authors have also constructed phenomenological Lagrangians in terms of hadron fields but driven by {\em ad hoc} assumptions whose link with QCD is not proven and which main goal is to simplify the structure of the theory. Well known examples of these models describing the strong interaction in the presence of resonances are the
Hidden Symmetry or Gauge Symmetry Lagrangians \cite{models} where vector mesons are introduced
as gauge bosons of suggested local symmetries. Most of these models lack naturalness and,
in a first approach, even consistency with QCD. However this last problem can be fixed
through a cautious repair \cite{rcht2}.
\vspace*{0.2cm} \\
\underline{\em Parameterisations} \vspace*{0.2cm} \\ \hspace*{0.3cm} Instead of resorting to a field theory another
possibility arises from the construction of dynamically driven parameterisations. The
main idea underlying this procedure is to provide an expression for the amplitudes 
as suggested by the supposed dynamics: resonance dominance, polology, etc. The usual simplicity
of these parameterisations looks very convenient for the analysis of experimental data though
the connection between the parameters and QCD is missing and, therefore, very little is 
understood about Nature with this approach. We will come back in the next Section to this 
widespread method.  
\vspace*{0.2cm} \\
\hspace*{0.3cm}
In order to explore the non-perturbative regime of the strong interaction theory are 
of special interest those semileptonic processes driven by the
hadronization of a QCD current into exclusive channels, like $e^{+} e^{-} \rightarrow H^{0}$ or
hadronic tau decays, due both to the fact that, on one
side, lepton and hadron sectors factorise cleanly and, in addition, exclusive channels give valuable information on the dynamics of the interaction itself, hence on the realization of 
non-perturbative QCD in this energy region.
\par
Within the Standard Model the matrix amplitude for the exclusive hadronic decays of the tau 
lepton, $\tau^{-} \rightarrow H^{-} \nu_{\tau}$, is generically given by
\beq
{\cal M} \, = \, \Frac{G_F}{\sqrt{2}} \,  
V_{\tiny{CKM}} \, \overline{u}_{\nu_{\tau}} \gamma^{\mu} 
\lef u_{\tau} \, {\cal H}_{\mu} \, , 
\label{eq:m}
\eeq
where
\beq
{\cal H}_{\mu}\,  = \, \left\langle \, H \,\left| \,\left( {\cal V}_{\mu} - {\cal A}_{\mu}\right) \,
e^{i {\cal L}_{QCD}} \,\right| \, 0 \, \right\rangle \; ,
\label{eq:h}
\eeq
is the hadron matrix element of the left current (notice that it has to be evaluated 
in the presence of the strong interactions driven by ${\cal L}_{QCD}$). Symmetries help
to define a decomposition of ${\cal H}_{\mu}$ in terms of the allowed Lorentz structures
of implied momenta and a set of functions of Lorentz invariants, the {\em hadron form
factors} $F_i$ of QCD currents, 
\beq
{\cal H}_{\mu} \, = \,  \,  \sum_i \! \! \underbrace{ \;   ( \,  \, \ldots \, \, )_{\mu}^i 
\; }_{Lorentz \, structure} \! \! \! \! \! \! 
 F_i (q^2, \ldots) \; .
\label{eq:ff}
\eeq
Therefore form factors are the specific goal to achieve and, as follows from the definition of ${\cal H}_{\mu}$ in Eq.~(\ref{eq:h}), arise 
from the evaluation of the relevant matrix elements of the vector and axial-vector currents
of QCD in its non-perturbative regime. It is significant to remark that these form factors
are universal and do not depend on the initial state, hence providing a compelling 
description of the hadronization of the QCD currents.
\par
An equivalent discussion of the hadronic decays of the tau lepton can be carried out
in terms of the {\em structure functions} $W_X$ defined in the
hadron rest frame \cite{KS2}~:
\beq
d\Gamma  =  \frac{G_F^2}{4 M_\tau} \, |V_{CKM}|^2 \, {\cal L}_{\mu\nu} \, {\cal H}^{\mu} \,
{\cal H}^{\nu*} \, dPS \;  ,
\eeq
with
\beq
{\cal L}_{\mu\nu} \, {\cal H}^{\mu} \, 
{\cal H}^{\nu*}  =  \sum_X \, L_X \, W_X \; ,
\label{eq:sff}
\eeq
where ${\cal H}_\mu$ is the hadronic current in Eq.~(\ref{eq:h}), ${\cal L}_{\mu\nu}$
carries the information of the lepton sector and $dPS$ collects the appropriate phase
space terms. Structure functions can be written in terms of the relevant form factors
and kinematical components. Accordingly they contain the dynamics of the hadronic decay and
their reconstruction can be accomplished through the study of spectral functions or 
angular distributions of data.
The number of structure functions depends,
clearly, of the number of hadrons in the final state. For a two-pseudoscalar case there
are 4 of them. For a three-pseudoscalar process the total number of structure functions
is 16.
\par
In this short review I will focus in the decays of the tau lepton into two and three
pions. There has been a lot of work in other channels but most of the dynamics,
modelizations and problems, to which I will pay attention, can be discussed in the pion case.
Moreover the theoretical description of the dynamics that drives the decays into
more than three pseudoscalars will have to wait until we understand well the three pion 
instance and, until present,
we can only rely in the model-independent isospin counting \cite{pppp}.

\section{Model building versus QCD}
In order to achieve a prediction or 
description of experimental data in any electroweak hadronic process, within the 
Standard Model, we need to input non-perturbative QCD into the analysis. If everything fits
usually we can determine parameters of the Standard Model. If something fails we can
claim a role for New Physics. In practice this kind of analysis is carried out with a modelization
of the strong interaction that is perhaps even inconsistent with QCD. As a consequence all
the procedure gets polluted or under suspicion.
\par
In the last years experiments like ALEPH, CLEO, DELPHI and OPAL 
\cite{exp0,exp1,exp15,exp2,exp21,exp24,exp25,exp3} have
collected an important, both because the amount and the quality, set of experimental data on hadronic decays of the tau lepton into exclusive channels. Lately the BABAR experiment is joining in this effort \cite{babar}.
Analyses of these data are carried out using the TAUOLA library \cite{tauola} that includes
parameterisations of the hadronic matrix elements. As we have emphasised above models
and {\em ad hoc} parameterisations may include simplifying assumptions not well controlled
from QCD itself. Therefore, while of importance to get an understanding of the dynamics
involved, they can be misleading and provide a delusive interpretation of data.
\par
In this Section we will describe the most relevant modelizations and the basis of the
effective field theory approach. Both of them are employed in the study of hadron decays
of the tau lepton.

\subsection{Breit-Wigner parameterisations}
The dynamics of a hadronic process in an energy region populated by resonances is mainly
driven by those states. This is the well known concept of {\em resonance dominance}
that has pervaded hadron dynamics since the first stages of the study of the strong 
interaction. Its application to the hadronization of charged QCD currents in tau decays has a 
long story \cite{preKS,KS1} that boils down into a series of papers
\cite{KS3,KS4} that carry an exhaustive analysis of the tau decays up to three pseudoescalars.
The CLEO collaboration has also applied this methodology to explore the resonance
structure in four pion decays \cite{exp3}.
\par
The parameterisation is accomplished by combining Breit-Wigner factors ($BW_R(q^2)$) according to 
the expected resonance dominance in each channel, for instance,
\begin{equation}
F(q^2) = {\cal N} \, \sum_i \, \alpha_i \, BW_{R_{i}}(q^2) \; ,
\end{equation} 
where $ {\cal N} $ is a normalisation and, in general, the expression is not necessarily linear
in the Breit-Wigner terms.  
Then data are analysed by fitting the $\alpha_i$ parameters and those present in the Breit-Wigner
factors (masses, on-shell widths). Two main models of parameterisations have been considered~:
\vspace*{0.2cm} \\
a) {\em K\"uhn-Santamar\'{\i}a Model (KS)} \\
The Breit-Wigner factors are given by \cite{preKS,KS1}
\begin{equation}
BW_{R_i}^{KS}(s) = \Frac{M_{R_i}^2}{M_{R_i}^2 -s-i \, \sqrt{s} \, \Gamma_{R_{i}}(s)}  \, ,
\label{eq:KS}
\end{equation} 
that guarantees the right asymptotic behaviour, ruled by QCD, for the form factors. 
\vspace*{0.2cm} \\
b) {\em Gounaris-Sakurai Model (GS)} \\
Originally constructed to study the role of the $\rho(770)$ resonance in the vector form
factor of the pion \cite{GS}, its use has been
extended to other hadronic resonances \cite{exp0,exp1,KS4}. The Breit-Wigner function
now reads~:
\begin{equation}
BW_{R_i}^{GS}(s) = \Frac{M_{R_i}^2+ f_{R_i}(0)}{M_{R_i}^2-s+f_{R_i}(s)-i \sqrt{s} \, \Gamma_{R_i}(s)} \, ,
\label{eq:GS}
\end{equation} 
where $f_{R_i}(s)$ carries information on the specific dynamics of the resonance 
and $f_{\rho(770)}(s)$ can be read from Ref.~\cite{GS}. 
\vspace*{0.3cm}\\
In both models the form factors are normalised in order to satisfy the chiral symmetry of 
massless QCD, at $s \ll M_{\rho}^2$, as ruled by the leading ${\cal O}(p^2)$ in the
$\chi PT$ expansion. 
The methodology applied by the experimental 
groups when using these parameterisations \cite{exp0,exp1} is to regard both models and consider the discrepancy between them as an estimate of the theoretical error.
\par
It is important to stress that the
simplicity of these parameterisations is obscured because the lack of a clear link
between them and QCD. They could even be at variance with the fundamental theory.

\subsection{Effective field theories and other model-independent knowledge}
Instead of relying on arguable parameterisations one can try to extract information about form factors
on grounds of S-matrix theory properties or QCD itself. The appealing aspect of this procedure
is that, when analysing data, the connection with the basic theory is generally clear and, therefore,
we can obtain a great deal of information on the hadronization procedure and, accordingly, on 
QCD itself. Here we will comment briefly on the hints to consider.
\par
On general grounds local causality of the interaction translates into the analyticity 
properties of the amplitudes and, correspondingly, of form factors. Being analytic functions
in complex variables the behaviour of form factors at different energy scales is related
and, moreover, they are completely determined by their singularities. Dispersion relations
embody rigorously these properties and are the appropriate tool to enforce them. In addition
unitarity must be satisfied in all physical regions. This S-matrix property provides precise
information on the relevant contributions to the spectral functions of correlators of 
hadronic currents that are closely related to form factors. Furthermore
a theorem put forward by S.~Weinberg \cite{chiral} and worked out by H.~Leutwyler \cite{leut} 
states that, if one writes down the most general possible Lagrangian, including all terms
consistent with assumed symmetry principles, and then calculates matrix elements with this
Lagrangian to any given order of perturbation theory, the result will be the most general
possible S-matrix consistent with analyticity, perturbative unitarity, cluster decomposition
and the principles of symmetry that have been specified.
\par
Besides, it has been pointed out \cite{ncc} that the inverse of the number of colours of the
gauge group $SU(N_C)$ could be taken as a perturbative expansion parameter. Indeed 
large-$N_C$ QCD shows features that resemble, both qualitatively and quantitatively, the
$N_C=3$ case. Relevant consequences of this approach are that meson dynamics in the 
large-$N_C$ limit is described by the tree diagrams of an effective local Lagrangian;
moreover, at the leading order, one has to include the contributions of the infinite
number of zero-width resonances that constitute the spectrum of the theory.
\par
It is on all these statements that part
of the model-independent work on low and intermediate energy hadronic dynamics has been 
based upon and in the following we resume the specific tools that this procedure has
set up.
\par
Massless QCD is symmetric under global independent $SU(N_F)$ rotations of left- and
right-handed quark fields
\beqn
q_L &\rightarrow& h_L \, q_L \; , \;\; \; \; \; \;h_L \in SU(N_F)_L \; , \nonumber \\ 
q_R &\rightarrow& h_R \, q_R \; , \;\; \; \; \; \;h_R \in SU(N_F)_R \; ,
\eeqn
where $N_F = 2,3$ is the number of light flavours. This is the well-known 
$SU(N_F)_L \otimes SU(N_F)_R$ chiral symmetry of QCD \cite{chiral,chiral1}. Quark masses
break explicitly this symmetry but what is more relevant is that it seems it is also
spontaneously broken. Though a rigorous prove of this feature has only been achieved
in the large number of colours limit \cite{NC}, the known phenomenology supports that
statement. Goldstone theorem demands the appearance of a phase of massless bosons associated
to the broken generators of the symmetry and their quantum numbers happen to correspond
to those of the lightest octet ($N_F = 3$) of pseudoscalars. Their non-vanishing masses 
are generated by the explicit breaking of chiral symmetry through quark masses. This 
Goldstone phase is timely because it provides an energy gap into the meson spectrum
between the octet of pseudoscalars and the heavier mesons starting with the $\rho (770)$.
We can take precisely the mass of this resonance, $M_\rho$, as a reference scale to 
introduce effective actions of QCD~: 
\vspace*{0.1cm} \\
a) \underline{\bf{$E \ll M_{\rho}$}}
\vspace*{0.2cm} \\
\hspace*{0.5cm} In this energy region chiral symmetry is the guiding
principle to follow. The relevant effective theory of QCD is {\em Chiral
Perturbation Theory} \cite{chiral,chiral1} that exploits properly
the chiral symmetry $SU(N_F)_L \otimes SU(N_F)_R$. In this effective 
action the active degrees of freedom are those of the octet of 
pseudoscalars and the heavier spectrum has been integrated out. As its
own name implies, $\chi$PT is a perturbation theory in the momenta of
pseudoscalars over a typical scale $\Lambda_{\chi} 
\sim 1 \, \mbox{GeV}$. This entails that the interaction vanishes with the 
momentum, giving an example of dual behaviour between the effective action
(perturbative at low energies) and QCD (where asymptotic freedom prevents
a perturbative expansion in that energy regime). By demanding that 
the interaction satisfies chiral symmetry the complete structure of the 
operators, at a definite perturbative order, is defined. However chiral
symmetry does not give any information on their couplings 
that, in general, carry the information of the contributions of heavier
states that have been integrated out \cite{vmd1,taron}. 
\vspace*{0.2cm} \\
b) \underline{\bf{$E \simeq M_{\rho}$}}
\vspace*{0.2cm} \\
\hspace*{0.5cm} At $E \simeq 1 \, \mbox{GeV}$ other meson states
are active degrees of freedom to take into account.
Chiral symmetry still provides the guide in the construction of the
effective action of QCD, following the pioneering work of S. Weinberg
\cite{chiral2} in which the new states are represented by fields that 
transform non--linearly under the axial part of the chiral group.
For the lightest octet of resonances (vectors, axial--vectors,
scalars and pseudoscalars) this procedure was carried out in Ref. 
\cite{vmd1} and the resulting Lagrangian is the basis of the {\em Resonance
Chiral Theory}. As in $\chi$PT, chiral symmetry constraints the
structure of the operators but gives no information on their couplings,
that remain unknown, though they could be studied either by using models or through
the analyses of Green Functions \cite{greenf}.
Therefore $R \chi T$ provides
a model--independent parameterisation
of the processes involving resonances and pseudoscalars in terms of 
those couplings. 
\par
Strong interactions in the resonance region lack an expansion
parameter that could provide a perturbative treatment of the amplitudes.
The large-$N_C$ limited pointed out above could yield the appropriate tool
for such an expansion and it is been employed as a guiding principle when
using $R \chi T$. The $1/N_C$ expansion tells us that, at leading order,
we should only consider the tree level diagrams given by a local Lagrangian
with infinite states of zero-width in the spectrum. $R \chi T$
provides the Lagrangian to be used at tree level but with a finite number
of zero-width resonances. However in most processes like hadron tau
decays, we need to include finite widths for the resonances. These only 
appear at next-to-leading order in the large-$N_C$ expansion that means
one-loop evaluations in $R \chi T$, aspect poorly known at present 
\cite{rchloop1,rchloop2}. Thus in practice we have to model this expansion to some extent and control
the relevance of the extra assumptions.
\vspace*{0.2cm} \\
c) \underline{\bf{$E \gg M_{\rho}$}}
\vspace*{0.2cm} \\
\hspace*{0.5cm} At much higher energies the asymptotic freedom of QCD
implies that a perturbative treatment of the theory is indeed appropriate. 
The study of this energy region is also important for low--energy
hadron physics because we are able to evaluate, within QCD, the
asymptotic behaviour of Green functions and then, through a matching 
operation, to impose these
constraints on the low--energy regime \cite{rcht2}. This heuristic procedure is well supported at
the phenomenological level \cite{greenf}. 
\vspace*{0.3cm} \\
To take advantage of the underlying QCD dynamics in the model--independent
study of hadronic observables, we have to consider the entire 
view and how different energy regimes intertwine between themselves.

\subsection{Hadronic off-shell widths of meson resonances}
The hadronic decays of the tau lepton happen in an energy region where resonances
do indeed resonate. Hence the leading large-$N_C$ prescription of zero-width resonances
has to be overcome. The introduction of finite widths, as seen in Eqs.~(\ref{eq:KS},\ref{eq:GS}), results in a new problem that needs
to be considered. For narrow resonances, like most of those with $I=0$ in the energy region
spanned by tau decays, it is a good approximation to consider constant widths that can be
taken from the phenomenology at hand. Wider resonances, though, have an off-shell structure
that has to be taken into account.  
\par
The off-shell width of the $\rho(770)$ has been studied thoroughly and it is dominated by 
the $\pi \pi$ contribution. In the KS or GS 
parameterisations the imaginary part of the mass in the pole reads \cite{GS}~:
\beq
\sqrt{s}  \, \Gamma_{\rho}(s)  =  \Gamma_{\rho}(M_{\rho}^2)   \Frac{s}{M_{\rho}} 
\frac{\sigma^3(s)}{\sigma^3(M_\rho^2)}  \, \theta \left( s-4m_{\pi}^2 \right)  ,
\label{eq:wGS}
\eeq
where $\sigma(s) = \sqrt{ 1-4m_{\pi}^{2}/s }$. 
In Ref.~\cite{widthro} if was seen that this width can be evaluated within $R \chi T$
through a Dyson-Schwinger like resummation controlled by the short-distance behaviour
required by QCD on the correlator of two vector currents. The result for the imaginary 
part of the pole is~:
\beq
M_{\rho} \, \Gamma_{\rho}(s) \, = \, \Frac{M_V^2 \, s}{96 \, \pi \, F^2} \,
 \sigma^3(s) \, \theta \left( s-4m_{\pi}^{2} \right)  \, ,
\label{eq:wUS}
\eeq
were $M_V$ is the vector resonance mass and $F$ the decay constant of the pion, both
of them in the chiral limit. 
It is worth to notice that the dependence on the $s$ variable of both imaginary parts
in Eqs.~(\ref{eq:wGS},\ref{eq:wUS})is the same.
\par
The hadronic off-shell widths for other resonances like $\rho(1450)$ or $a_1(1260)$, that
are also relevant in the decays of the tau lepton, is not so well known. In principle
the methodology put forward in Ref.~\cite{widthro} could also be applied but it is necessary
to know better the perturbative loop expansion of $R \chi T$ in order to proceed. Therefore
one has to resort to appropriate modelizations and the key point is, of course, the leading
behaviour of the off-shell structure in the $s$ variable. Hence it is customary to 
propose parameterised widths of which the simplest version reads as~:
\beq
\Gamma_R(s) = \Gamma_R(M_R^2) \Frac{\Phi(s)}{\Phi(M_R^2)} \left( \Frac{M_R^2}{s} \right)^{\alpha} 
\theta(s-s_{th}),
\label{eq:wa1}
\eeq
where the $\Phi(s)$ function is related with the available phase space that corresponds to 
the threshold given by $s_{th}$. The $\alpha$ parameter can be given by models or fitted to
the experimental data. This last procedure was used in Ref.~\cite{us3pion} to obtain information
on the $a_1(1260)$ width from the tau decay into three pions, giving $\alpha \sim 5/2$. A 
thorough study on the off-shell widths of resonances that QCD demands is still missing.

\section{$\tau^- \rightarrow \pi^- \pi^0 \nu_{\tau}$~: The vector form factor of the pion}
The vector form factor of the pion, $F_V(s)$, is defined through~:
\beq
\left\langle  \pi^+(p') \, \pi^-(p)  \left| \, {\cal V}_{\mu}^{3} \, \right| 
 0  \right\rangle 
 =   \, \left( p - p' \right)_\mu F_V(s)   ,
\eeq
where $s=(p+p')^2$ and ${\cal V}_{\mu}^3$, the third component of the vector current
associated to the $SU(3)$ flavour symmetry of the QCD Lagrangian.  This form factor drives
the isovector hadronic part of $e^+ e^- \rightarrow \pi^+ \pi^-$ and, in the isospin limit,
of $\tau^- \rightarrow \pi^- \pi^0 \nu_{\tau}$.
At very low energies, $E \ll M_{\rho}$, $F_V(s)$ has been studied in the $\chi PT$ framework up to 
${\cal O}(p^6)$ \cite{op6fv,cfu}. The analysis of this form factor in the resonance energy region
has a long history in the literature where a great deal of procedures have been applied though
here we will collect the last developments.
\vspace*{0.2cm} \\
a) \underline{$M_{\rho} \lsim E \lsim 1 \gev$}
\vspace*{0.1cm} \\
This energy region is dominated by the $\rho(770)$ and, accordingly, its study is
relevant to determine the parameters of this resonance. In addition it gives the largest
contribution to the hadronic vacuum polarisation piece of the anomalous magnetic moment
of the pion \cite{g-2}.
\par
The authors of Ref.~\cite{gp97}
proposed a framework where the ${\cal O}(p^4)$ $\chi PT$ low-energy result is matched
 at higher energies with an expression driven by vector meson dominance that is modulated by an
Omn\`es solution of the dispersion relation satisfied by the vector form factor
of the pion. It provides an excellent description of the $\rho(770)$ up to energies of $1 \gev$.
The more involved procedure of the unitarization approach \cite{oop} gives also a good description
of this energy region.
\par
A model-independent parameterisation of the vector form factor constructed on grounds of 
an Omn\`es solution for the dispersion relation has also been considered \cite{pp01,pp02,ty}. 
This approach can be combined with $R \chi T$ \cite{pp01} and it is able to give some improvement over the previous approach if one includes
information on the $\rho(1450)$ through the $\pi \pi$ elastic phase-shift input in the 
Omn\`es solution. Hence it extends the description of the form factor up to $E \sim 1.3 \gev$.
\par
\begin{figure}[t]
\begin{center}
\hspace*{-0.3cm}\includegraphics*[scale=0.56]{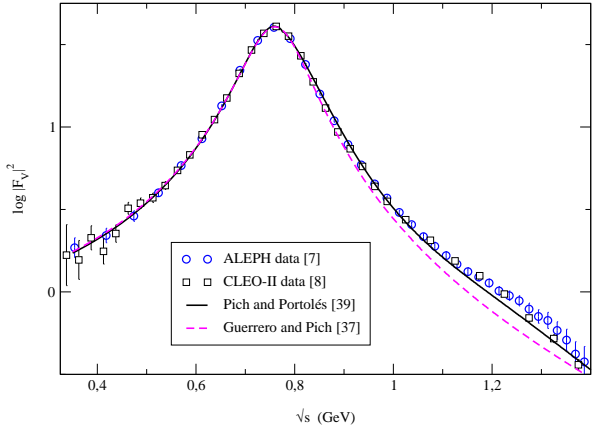}
\vspace*{-0.9cm}
\caption[]{\label{fig:1} Comparison of the vector form factor of the pion as given by
tau data and the theoretical description of Refs. \cite{gp97,pp01}.}
\end{center}
\end{figure}

A comparison of the theoretical descriptions given by Refs.~\cite{gp97,pp01} and the 
experimental data by ALEPH \cite{exp0} and CLEO \cite{exp1} is shown in Fig.~\ref{fig:1}.

\vspace*{0.3cm}
\noindent
b) \underline{$1 \gev \lsim E \lsim 2 \gev$}
\vspace*{0.1cm} \\
The extension of the description of the vector form factor of the pion at higher energies
is cumbersome. Up to $2 \gev$ two $\rho-$like resonances play the main role~: $\rho(1450)$ and 
$\rho(1700)$. However the interference between resonances, the possible presence of a 
continuum component, etc. still deserve a study not yet done.
\par
\begin{figure}[t]
\begin{center}
\vspace*{-0.2cm}
\includegraphics*[scale=0.48]{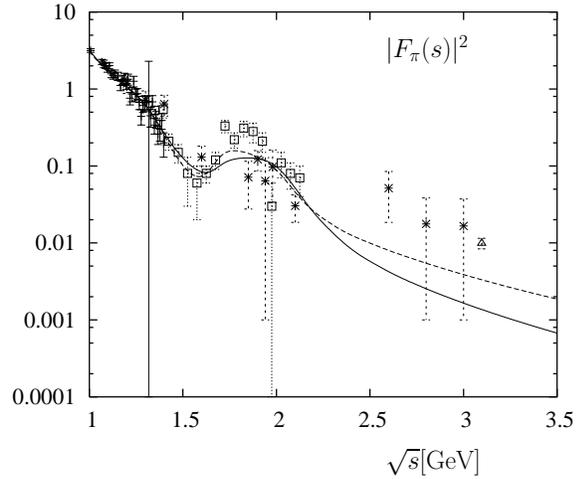}
\vspace*{-1.2cm}
\caption[]{\label{fig:2} Comparison of the vector form factor of the pion in the energy
region above $1 \gev$ from Ref.~\cite{KS4}. Solid (dashed) line corresponds to the KS
(GS) parameterisation. See that reference for the explanation of data.}
\vspace*{-0.8cm}
\end{center}
\end{figure}

The inclusion of $\rho(1450)$ only improves
slightly the behaviour when a Dyson-Schwinger-like resummation is performed in the
framework of $R \chi T$ \cite{rchloop1}.
 Lately, and based in a previous modelization proposed in 
Ref.~\cite{do01}, a procedure to extend the description of the vector form factor of 
the pion at higher energies has been put forward \cite{KS4}. The proposal for the 
form factor embodies
a Breit-Wigner parameterisation using both KS (\ref{eq:KS}) and GS (\ref{eq:GS}) models
to describe $\rho(770)$, $\rho(1450)$ and $\rho(1700)$ resonances, appended with a
modelization of large-$N_C$ QCD which sums up an infinite number of zero-width 
resonances to yield a Veneziano type of structure. In Fig.~\ref{fig:2} it is shown how
this parameterisation compares with data. The description is reasonable up to $2 \gev$.
Above this region there is almost no data though, in principle, it looks quite compatible
with it \cite{KUTAU}.

\vspace*{0.4cm} 
The all-important role that plays the vector form factor of the pion in the hadronic 
vacuum polarisation contribution to the anomalous magnetic moment of the pion \cite{g-2},
together with the seeming discrepancy between the predictions provided by 
$e^+ e^- \rightarrow \pi^+ \pi^-$ \cite{cmd2,kloe} and $\tau^- \rightarrow \pi^- \pi^0 \nu_{\tau}$
data, set up the issue of the size of isospin violation. A few years ago it was suggested
\cite{Davier} that the late excellent experimental determinations of the vector spectral 
functions in tau decay could be used to determine the hadronic vacuum polarisation 
contribution to $a_{\mu}$. The discrepancies that have arisen between the predictions from
both sources were first blamed to the radiative corrections analyses in the first CMD-2 data, then
to noticeable isospin violation unaccounted-for. A thorough analysis of the radiative
corrections in $\tau^- \rightarrow \pi^- \pi^0 \nu_{\tau}$ and other relevant isospin
violating sources (kinematics, short-distance electroweak corrections, $\rho-\omega$ mixing)
was carried out in Ref.~\cite{Vincenzo}. As the variance persisted it was also claimed
a strong isospin violation in the $\rho$ mass \cite{Jeger} though nor theory \cite{BG96}
nor other several determinations from experimental data \cite{pp02,ty} seem to support it. 
Further analysis of the CMD-2 procedure detected several mistakes that have been corrected 
and are, at present, in 
agreement with recent KLOE data \cite{kloe}. The matter is still open (see Ref.~\cite{Massimo} for a recent account of this problem).

\section{$\tau^- \rightarrow (\pi \pi \pi)^- \nu_{\tau}$~: Axial-vector form factors}
The hadronic matrix element that drives the decay of the tau lepton into three pions is parameterised by four form factors $F_i$ defined as~:
\begin{eqnarray} \label{eq:fff}
\left\langle \pi^{-0}(p_1) \pi^{-0}(p_2) \pi^{+-}(p_3) \left| \, \left( {\cal V}^{\mu} 
- {\cal A}^{\mu} \right) \, \right| 0 \right\rangle \, = \nonumber \\ \\  
V^{\mu}_1  F_1^A(Q^2,s_1,s_2)\,  + \, V^{\mu}_2 F_2^A(Q^2,s_1,s_2) \nonumber \\ \nonumber \\ 
+ \, Q^{\mu} \, F_3^A(Q^2,s_1,s_2) \, + \,  i \, V^{\mu}_3 F_4^V(Q^2,s_1,s_2) \, , \nonumber
\end{eqnarray} 
where
\begin{eqnarray}
V^{\mu}_1 &=& \left( g^{\mu\nu} - \frac{Q^{\mu}Q^{\nu}}{Q^2}\right)  (p_1-p_3)_{\nu} \, ,\nonumber \\
V^{\mu}_2 &=& \left( g^{\mu\nu} - \frac{Q^{\mu}Q^{\nu}}{Q^2}\right)  (p_2-p_3)_{\nu} \, ,\nonumber \\
V^{\mu}_3 & = & \varepsilon^{\mu\alpha\beta\gamma} p_{1 \alpha} p_{2\beta} p_{3\gamma} \, ,\nonumber \\
Q^{\mu} & = & p_1^{\mu} + p_2^{\mu} + p_3^{\mu} \, , \\
s_i & = & \left( Q-p_i \right)^2 \nonumber \; .
\end{eqnarray}
This parameterisation is general for any three pseudoscalar final state. In the particular case
at hand (three pions), we have, due to Bose-Einstein symmetry, that $F_2^A(Q^2,s_1,s_2) = F_1^A(Q^2,s_2,s_1)$. The scalar form factor $F_3^A$ vanishes with the mass of the Goldstone
boson (chiral limit) and, accordingly, gives a tiny contribution in the three pion case.
Finally the vector current only contributes if isospin symmetry is broken as demanded by
G parity conservation; hence in the isospin limit $F_4^V=0$. 
\par
The final hadron system in the $\tau \rightarrow \pi \pi \pi \nu_{\tau}$ decays spans
a wide energy region $3 m_{\pi} \leq E \lsim M_{\tau}$ that is heavily populated by
resonances and has been thoroughly studied. In the very low energy regime the chiral
constraints where explored in the seminal Ref.~\cite{FW80} and lately it has been 
calculated up to ${\cal O}(p^4)$ in $\chi PT$ \cite{cfu}. To account for the resonance
energy region several modelizations given by phenomenological Lagrangians and vector
meson dominance, matching the chiral behaviour, have been 
employed \cite{cfu,pheno3}. However the most successful 
parameterisation has been the one given by the KS model \cite{preKS,KS1} that has been 
extendend to the study of all possible channels of three pseudoscalar mesons in the
final state \cite{KS3} and included in the TAUOLA Library \cite{tauola}.
\par
The dynamics of $\tau \rightarrow \pi\pi\pi\nu_{\tau}$ is driven by the presence of the
axial-vector $a_1(1260)$ modulated by the $I=1$ vector resonances $\rho(770) (\rho)$,
$\rho(1450) (\rho')$ and $\rho(1700) (\rho'')$. Hence in the KS model the spin 1 axial-vector
form factor is given by~:
\begin{eqnarray}
F_1^A(Q^2, s_i) = {\cal N}|_{\chi {\cal O}(p^2)} \, BW_{a_1}(Q^2) \, \times \nonumber \\ \\
\times \; 
\Frac{BW_{\rho}(s_i) + \alpha \, BW_{\rho'}(s_i) \, + \, \beta \, 
 BW_{\rho''} (s_i)}{1+\alpha+\beta} \; .
\nonumber
\end{eqnarray}
This description, complemented with an {\em ad hoc} construction of the off-shell width 
of the $a_1(1260)$ resonance, provides a good description of the spectrum of three pions
\cite{KS1} though a slight discrepancy shows up in the integrated structure functions \cite{exp24}.
The fit to the data gives the values of the $\alpha$ and $\beta$ parameters,
that compute the weight of each $\rho$-like resonance and, in addition, one can study
masses and on-shell widths of the participating resonances. The issue of isospin violation
in this channel, within the KS model, has also been considered \cite{isocp}. 
Lately it was shown that
this Breit-Wigner parameterisation is not consistent with chiral symmetry at ${\cal O}(p^4)$
and thus with QCD \cite{us3pion,joVictoria}.
\par
\begin{figure}
\hspace*{-0.5cm}
\includegraphics*[angle=0,scale=0.55,clip]{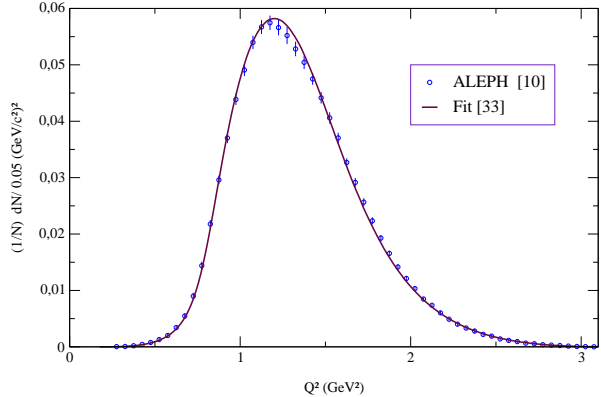}
\vspace*{-0.8cm}
\caption{\label{fig:3} Fit to the ALEPH data \cite{exp15} for the normalised
$\tau^- \rightarrow \pi^+ \pi^- \pi^- \nu_{\tau}$ decay \cite{us3pion}. }
\vspace*{-0.4cm}
\end{figure}

A thorough study of the 
axial-vector form factors in $\tau \rightarrow \pi\pi\pi\nu_{\tau}$ has been performed in 
Ref.~\cite{us3pion} using the methodology of effective field theories. The main components
of this approach are~:
\vspace*{0.2cm} \\
1/ {\em Resonance Chiral Theory}
\vspace*{0.1cm} \\
A Lagrangian theory provided by $R \chi T$ \cite{vmd1} has been employed. In this reference
only linear terms in the resonances, whose couplings are under control, were introduced.
In the decay of the tau into three
pions, as commented above, a basic role is played by the interplay between the $a_1(1260)$
and $\rho$-like resonances. Accordingly it was necessary to construct, following chiral symmetry,
a model-independent non-linear $a_1 \rho \pi$ coupling that introduced 5 unknown couplings.
\vspace*{0.2cm} \\
2/ {\em Modelization of Large-$N_C$}
\vspace*{0.1cm} \\
Following the ideas of large-$N_C$ there were considered the tree level diagrams of the local Lagrangian
theory. However this approach was corrected~: first by cutting the infinite number of 
resonances that the leading $1/N_C$ expansion asks for, considering one octet of resonances only;
then including off-shell widths for the $\rho(770)$ and $a_1(1260)$ according to the expressions
given above, Eqs.~(\ref{eq:wUS},\ref{eq:wa1}), into the resonance propagators.
\vspace*{0.2cm} \\
3/ {\em Asymptotic behaviour of form factors}
\vspace*{0.1cm} \\
The high energy behaviour of the axial-vector spectral function at leading 
order in perturbative QCD \cite{Eduard} demands, heuristically, that the axial-vector form
factors vanish at high $Q^2$, conclusion already collected into the Lepage-Brodsky leading
behaviour of form factors \cite{Brodsky}. 
This idea was used in order to provide two constraints on 
the 5 unknown couplings of the Lagrangian theory. Afterwards the evaluation of the Feynman 
diagrams constributing to $\tau \rightarrow \pi\pi\pi\nu_{\tau}$ showed that, once
the constraints are enforced, only one combination of couplings is left unknown.
\vspace*{0.35cm} \\
\hspace*{0.3cm}
Hence the authors get a parameterisation of the three pion decay of the tau lepton in terms
of four free parameters~: $M_{a_1}$, $\Gamma_{a_1}(M_{a_1})$, the combination of coupling
constants and the $\alpha$ parameter in the off-shell width of the $a_1(1260)$ resonance. Next
an analysis of the ALEPH data \cite{exp15} on the spectrum and branching ratio
of $\tau^- \rightarrow \pi^+ \pi^- \pi^- \nu_{\tau}$ is performed. 
The fit is shown in Fig.~\ref{fig:3} and they obtain $M_{a_1} = (1.204 \pm 0.007) \gev$
and $\Gamma_{a_1}(M_{a_1}^2) = (0.48 \pm 0.02) \gev$ where the errors, given by the 
minimisation program, are only statistical.

\begin{figure*}[!t]
\hspace*{0.5cm}
\includegraphics*[angle=-90,scale=0.66,clip]{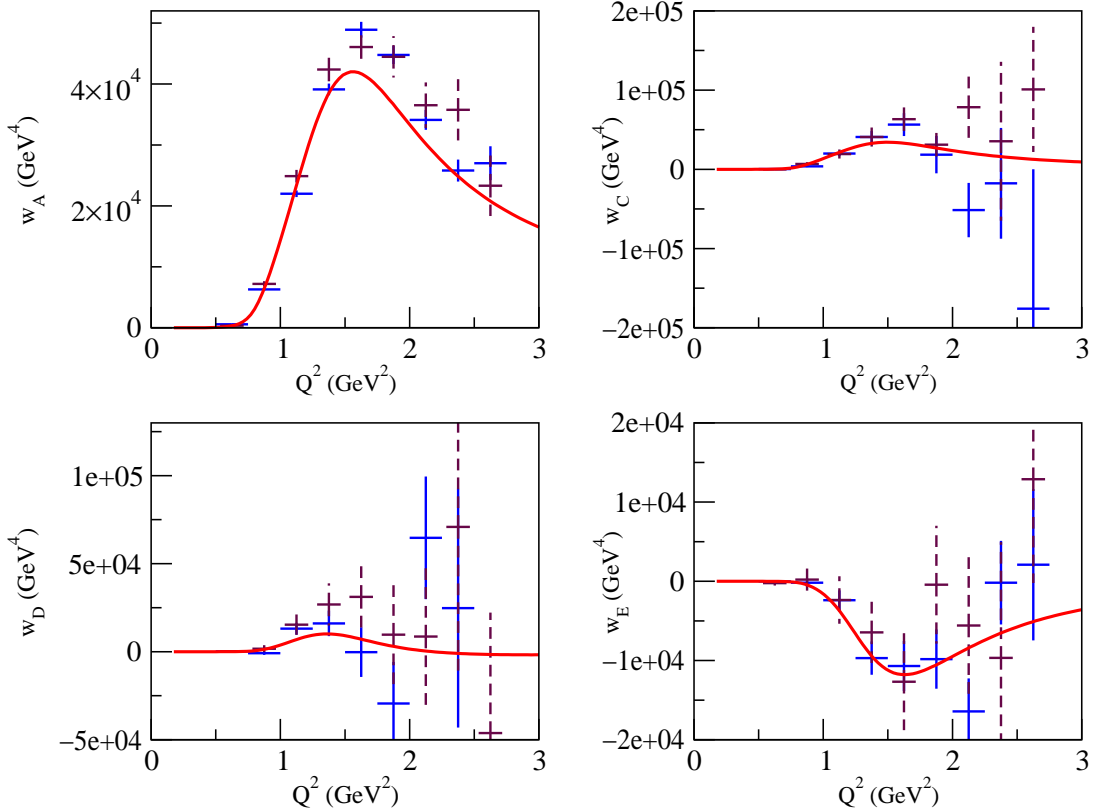}
\vspace*{-0.4cm}
\caption{\label{fig:4} Theoretical predictions for the $w_A$, $w_C$, $w_D$
and $w_E$ integrated structure functions \cite{us3pion} in comparison with the experimental
data from CLEO-II (solid) \cite{exp24} and OPAL (dashed) \cite{exp2}.}
\vspace*{-0.3cm}
\end{figure*}
OPAL \cite{exp2} and CLEO \cite{exp24} have
collected data on the dominant structure functions in the 
$\tau^- \rightarrow \pi^- \pi^0 \pi^0 \nu_{\tau}$ decay, namely, $W_A$, $W_C$,
$W_D$ and $W_E$ (\ref{eq:sff}) that drive the contribution of the $J=1^+$ amplitude
 into the process and,
accordingly, the {\em integrated structure 
functions} over all the available phase space, defined as~:
\beq
w_{A,C} \, = \, \int \, ds_1 \, ds_2 \, W_{A,C}  \; ,
\eeq
\beq
w_{D,E} \, = \, \int \, ds_1 \, ds_2 \, sign\left( s_1 - s_2 \right) \, W_{D,E} \; .  
\eeq
CLEO \cite{exp24} displays the forecast given by the KS model and notice a slight 
discrepancy that shows up mainly in $w_A$. Then in order to have a better description
they modify the model by supplying 
some quantum-mechanical structure (a heritage of nuclear physics that accounts for the
finite size of hadrons) \cite{modcleo} that yields a good fit to data.
\par
Following the results of the effective field theory approach explained above, and once the
parameters are determined, it is possible to predict
the integrated structure functions.  By 
assuming isospin symmetry one can use the information obtained from the 
charged pions case to provide a description for the $\pi^- \pi^0 \pi^0$ final hadronic
state. The result and its comparison with the data is shown in Fig.~\ref{fig:4}. For
$w_C$, $w_D$ and $w_E$, it can be seen that there is a good agreement in the low 
$Q^2$ region, while for increasing energy the experimental errors become too large
to state any conclusion (moreover, there seems to be a little disagreement between
both experiments at several points). On the other hand, in the case of $w_A$ the 
theoretical curve seems to lie somewhat below the data for $Q^2 \gsim 1.5 \gev^2$.
However the study carried out in Ref.~\cite{us3pion} seems to conclude that this is 
due to some inconsistency between the data by CLEO and OPAL, on one side, and ALEPH
on the other. If the present data for $w_A$ is confirmed and the errors in the
region of discrepancy lessen we will face a new structure not accounted for in the 
present theoretical description.

\vspace*{-0.1cm}
\section{Prospects}
Recently the CLEO Collaboration has published an analysis of the data collected on the 
$\tau^- \rightarrow K^+ K^- \pi^- \nu_{\tau}$ decay \cite{KLEO}. It was 
known that this process is not well described by the KS model \cite{cleoprob} and therefore
in the new analysis they have reshaped the model with two new arbitrary parameters that
modulate both 
one of the axial-vector and the vector form factors (\ref{eq:fff}).
Afterwards all the parameters are obtained through a reasonable fitting procedure.
Along these pages we have emphasised the fact that arbitrary parameterisations are of
little help in the procedure of obtaining information about non-perturbative QCD. It
is usually claimed that the parameters hide physics that is not included specifically
 in the theory and
they measure in fact our ignorance~: the conclusion is that
we may know how much ignorant we are but do not have a clue about the way out. In the CLEO example
just pointed out the new parameter in the vector form factor spoils the Wess-Zumino
anomaly normalisation, that appears at ${\cal O}(p^4)$ in $\chi PT$. It is true that there
are non-anomalous contributions proportional to the pseudoscalar masses at the next perturbative
order that could
account for a deviation but it would be surprising that the correction is around $80 \%$
as the fit points out. The real issue is that, as we have indicated, the KS model is not
consistent with QCD and the CLEO reshaping is of not much use.
\par
Hadron decays of the tau lepton can provide all-important information on the hadronization
of currents in order to yield relevant knowledge on non-perturbative features of 
low-energy Quantum Chromodynamics. In order to achieve this goal we need to input more 
controlled QCD-based
modelizations. Our target is not only to fit the data at whatever cost, 
but do it with a reasonable parameterisation that allows us to understand more about the
theoretical description of Nature.
\par
The effective theory approach seems, along this line, more promising than the 
Breit-Wigner parameterisations. The procedure
relies in a field theory construction that embodies, up to a supposedly minor modelization of the 
large-$N_C$ behaviour, the relevant features of QCD in the resonance energy region, giving 
an appropriate account of the main traits of the experimental data and showing
that it is a compelling framework to work with. Notwithstanding it still has to 
probe its case with a complete study of, at least, all the three pseudoscalar channels.
\par
Hadronic tau decays have undergone, during the last years, a fruitful era of excellence
from the point of view of collecting experimental data. Experimentalists have done and are
doing a great job. Now time has come for theoreticians to do their task.
\vspace*{-0.2cm}
\section*{Acknowledgements}
I wish to thank the organisers of the TAU04 meeting in Nara (Japan) for their excellent job.
This work has been supported in part by MCYT (Spain) under Grant FPA2001-3031, by Generalitat
Valenciana (Grants GRUPOS03/013 and GV04B-594) and by EU HPRN-CT2002-00311 (EURIDICE).

\end{document}